\documentclass{aa}  

\usepackage{graphicx}
\usepackage{epstopdf}
\usepackage{txfonts}
\begin{document}

   \title{Dust composition and mass-loss return from the luminous blue variable R71 in the LMC}                                                                                                                                                                                                                                                                                


   \author{S. Guha Niyogi
          \inst{1}
          \and
          M. Min
          \inst{2}
          \and
          M. Meixner
          \inst{3}
          \and
          L.B.F.M. Waters
          \inst{4}
          \and
          J. Seale
          \inst{3}
          \and
          A.G.G.M. Tielens
          \inst{1}}

   \institute{Leiden Observatory, Leiden University, PO Box 9513, NL-2300RA Leiden, The Netherlands\\
              \email{suklima@strw.leidenuniv.nl}\\
         \and
             Astronomical Institute ``Anton Pannekoek'', University of Amsterdam,\\ 
                     PO Box 94249, 1090 GE, Amsterdam, The Netherlands\\     
         \and
             Space Telescope Science Institute, 3700 San Martin Drive, Baltimore, MD 21218, USA\\
         \and
             SRON, Sorbonnelaan 2, 3584 CA, Utrecht, The Netherlands\\}
             

   \date{Received 3 March 2014 /Accepted 29 July 2014}

 
  \abstract{
   
   \textit{Context.} {We present an analysis of mid-
   and far-infrared (IR) spectrum 
   and spectral energy distribution (SED) of the luminous blue variable 
   (LBV) R71 in the Large Magellanic Cloud (LMC).}\\
   \textit{Aims.} {This work aims to understand the overall contribution of high-mass LBVs to the 
   total dust-mass budget of the interstellar medium (ISM) of the LMC and compare this with the contribution from 
   low-mass asymptotic giant branch (AGB) stars. As a case study, we analyze the SED of R71.} \\
   \textit{Methods.} {We compiled all the available 
   photometric and spectroscopic observational fluxes from various telescopes for a wide 
   wavelength range (0.36 -- 250\,$\mu$m). We determined the dust composition from the 
   spectroscopic data, and derived the ejected dust mass, dust mass-loss rate, and other dust shell 
   properties by modeling the SED of R71. We noted nine spectral features in the dust shell of R71 by analyzing 
   \textit{Spitzer Space Telescope} spectroscopic data. Among these, we identified
   three new crystalline silicate features. We computed our model spectrum by using 3D radiative transfer 
   code MCMax.} \\
   \textit{Results.} {Our model calculation shows that dust is dominated by 
   amorphous silicates, with some crystalline silicates, metallic iron, and a very tiny 
   amount of polycyclic aromatic hydrocarbon 
   (PAH) molecules. The presence of both silicates and PAHs indicates that the dust has a mixed chemistry. 
   We derived a dust mass of 0.01 M$_\odot$, from which we arrive at a  
   total ejected mass of $\approx$ 5 M$_\odot$. This implies 
   a time-averaged dust mass-loss rate of 2.5$\times$10$^{-6}$ M$_\odot$\,yr$^{-1}$ 
   with an explosion about 4000 years ago. We assume that
   the other five confirmed dusty LBVs in the LMC loose mass at a similar rate,
   and estimate the total contribution to the mass budget of the LMC to be
   $\approx$ 10$^{-5}$ M$_\odot$\,yr$^{-1}$,
   which is comparable to the contribution by all the AGB stars in the LMC.}\\
   \textit{Conclusions.} {Based on our analysis on R71, we speculate that LBVs as a class 
   may be an important dust source in the ISM of the LMC.}

   }

   \maketitle

\section{Introduction}
  \label{Intro}
  
  Dust is a key component of the interstellar medium (ISM) of galaxies. Grains form primarily 
  in the ejecta of the stars. The two main sources of dust are the winds of low-to intermediate-mass 
  (M $<$ 8 M$_\odot$) stars and the rapidly evolving, high-mass stars (M $>$ 8 M$_\odot$). 
  In our Milky Way Galaxy (MWG), low-mass stars mainly dominate the injection of dust 
  into the ISM during the asymptotic giant branch (AGB) phase (Gehrz \cite{Gehrz-1989}). 
  But the origin of dust in galaxies in the very early universe is controversial. 
  The observed high abundance of dust in galaxies in the early Universe has often been taken to 
  imply that rapidly evolving, high-mass stars are an important dust source, 
  although studies have shown that low-mass AGB stars can evolve fast enough in low-metallicity 
  environments (e.g., Vassiliadis \& Wood \cite{Vassiliadis-Wood-1993}; 
  Marigo \cite{Marigo-2000}; Sloan et al. \cite{Sloan-et-al-2009}) 
  to contribute substantial amounts of dust.

  The evolution of high-mass stars in galaxies is dominated
  by high mass-loss throughout their entire lives. Depending on the initial mass
  of the stars, their mass-loss rates vary during their evolution, and this in turn
  determines their final fates. The initial mass and mass-loss rate of a star determines
  whether it will end its life  as a red supergiant (RSG) 
  and explode as a supernovae (SNe), or evolve through a blue supergiant (BSG) 
  to an unstable luminous blue variable (LBV) phase and subsequently move to 
  the Wolf-Rayet (WR) phase before
  exploding as a SNe. Specifically, according to the stellar evolutionary scenario
  (e.g., Maeder \cite{Maeder-1983}; Chiosi \& Maeder \cite{Chiosi-Maeder-1986}; 
  Meynet et al. \cite{Meynet-et-al-2011}), 
  a star with an initial mass M$_{ZAMS}$ $\geq$ 40 M$_{\odot}$ undergoes the 
  following evolutionary path:
  
  $O \rightarrow BSG \rightarrow LBV \rightarrow WR \rightarrow SN$.
  
  After a star has exhausted all of its hydrogen in its core, the core starts to shrink.
  This causes hydrogen-shell burning, which leads to a rapid redward evolution at more or 
  less constant luminosity. But at some point the star becomes unstable, and mass loss
  increases dramatically. This phase of instability is known as the LBV phase. 
  The LBVs show photometric variability ranging from giant eruptions of $>$ 2 magnitude
  to small oscillations of $\simeq$ 0.1 magnitude 
  (Lamers \& Fitzpatrick \cite{Lamers-Fitzpatrick-1988}). These stars are intrinsically 
  very bright with a luminosity of $\simeq$ 10$^6$ L$_{\odot}$ (Humphreys \& Davidson 
  \cite{Humphreys-Davidson-1984}).
  Eventually, the star will strip off its outer hydrogen-rich layers
  and become a WR star with a helium-rich atmosphere.

  For a star with initial mass between 30 M$_{\odot}$ $\leq$ M$_{ZAMS}$ $\leq$ 40 M$_{\odot}$, 
  the evolutionary path is almost the same, except that it 
  can possibly (but not necessarily) go through an additional RSG phase:

  $O \rightarrow BSG \rightarrow (RSG) \rightarrow  LBV \rightarrow WR \rightarrow SN$.


  
  One of the fundamental question in astronomy is how much dust is
  contributed to the total mass budget of a galaxy by these 
  high-mass stars (RSGs, LBVs, WRs, and SNe)? 
  For the RSGs, it has been predicted that the stars with
  initial mass M$_{ZAMS}$ $>$ 30 M$_{\odot}$ return approximately
  3--10 M$_{\odot}$ into the ISM (Jura \& Kleinmann \cite{Jura-Kleinmann-1990}).
  Even more mass can be ejected during the final SN explosion.
  Circumstellar dust around an LBV ($\eta$ Car) was first discovered by
  Westphal \& Neugebauer (\cite{Westphal-Neugebauer-1969}) in 
  the MWG. 
  During this 
  phase, the atmosphere is very loosely bound to the star and any small disturbance 
  may lead to a high mass-loss rate typically between
  $10^{-6} < \dot{M} < 10^{-4}$\,M$_\odot$\,yr$^{-1}$ 
  (e.g., Stahl \cite{Stahl-1987}; Lamers \& Fitzpatrick \cite{Lamers-Fitzpatrick-1988};
  Lamers \cite{Lamers-1997}) with an expansion velocity ($v_{exp}$)
  up to a few hundred km/s (Stahl \& Wolf \cite{Stahl-Wolf-1986}). 
  Over the average life-time of the LBVs ($\simeq10^5$ years), 
  they may shed around 1 M$_{\odot}$ or more through steady mass-loss and more during 
  major eruptions (e.g., Humphreys \& Davidson \cite{Humphreys-Davidson-1984}; 
  Stahl \cite{Stahl-1987}; Nota \& Lamers \cite{Nota-Lamers-1997}). 
  Similarly to the LBVs, the late-type WR stars (WC7 -- WC9) lose mass at high rates of
  $10^{-5} < \dot{M} < 10^{-4}$\,M$_\odot$\,yr$^{-1}$ 
  (e.g., Williams et al. \cite{Williams-et-al-1987};
  Crowther \& Smith \cite{Crowther-Smith-1997}),
  although in the MWG it has been estimated that late-type 
  WC stars contribute only little to the dust-mass 
  budget of the ISM (van der Hucht \cite{van-der-Hucht-et-al-1986}).

  It has long been assumed that most interstellar dust is 
  injected into the ISM of galaxies by
  low-mass stars during the 
  AGB phase. 
  During this phase, these stars lose mass at a 
  rate of $10^{-8} < \dot{M} < 10^{-4}$\,M$_\odot$\,yr$^{-1}$ 
  (vanLoon et al. \cite{vanLoon-et-al-2005}).
  Recently, it has become more and more apparent that high-mass stars play an important role 
  in the dust budget of the ISM of galaxies. 
  First, various authors (e.g., Omont et al. \cite{Omont-et-al-2001}; Bertoldi \& Cox \cite{Bertoldi-Cox-2002}; 
  Maiolino \cite{Maiolino-2006}; Dwek et al. \cite{Dwek-et-al-2007}; Gall et al. \cite{Gall-et-al-2011})
  have speculated that high-mass stars may be an important source for the dust budget of the galaxies at high redshifts 
  (as far back as 800 million years after the Big Bang). These authors have argued that at this early time, 
  low-mass stars have not yet had the time to evolve into the AGB phase, although other studies 
  have shown that low-mass AGB stars can evolve fast enough in low-metallicity 
  environments to produce the dust seen in the galaxies at high redshifts
  (e.g., Vassiliadis \& Wood \cite{Vassiliadis-Wood-1993}; 
  Marigo \cite{Marigo-2000}; Sloan et al. \cite{Sloan-et-al-2009}).

  Secondly, recent observations by \textit{Herschel Space Observatory} 
  (Pilbratt et al. \cite{Pilbratt-et-al-2010}) have revealed 
   $\simeq0.1$ M$_{\odot}$ of cold dust in the Galactic supernova remnant, Cas A, which corresponds to
  about 10\% of the condensible elements (Barlow et al. \cite{Barlow-et-al-2010}). 
  Because this emission originates in the region interior to the reverse shock, it must correspond 
  to dust condensed in the supernova ejecta. Similarly, a recent study has revealed that
  SN 1987A in the Large Magellanic Cloud (LMC) has produced $\simeq 0.5$ M$_{\odot}$ of dust 
  (Matsuura et al. \cite{Matsuura-et-al-2011}).
  Finally, various authors (e.g., Srinivasan et al. \cite{Srinivasan-et-al-2009}; 
  Matsuura et al. \cite{Matsuura-et-al-2009}; Riebel et al. \cite{Riebel-et-al-2012}; 
  Boyer et al. \cite{Boyer-et-al-2012})
  estimated the dust budget of the
  Magellanic Clouds. Their studies on the measured mass-injection
  rates by stars have focused almost exclusively on the AGB and RSG stars
  (e.g., Sargent et al. \cite{Sargent-et-al-2010}; \cite{Sargent-et-al-2011}; Riebel et al. \cite{Riebel-et-al-2012}). 
  Comparison of these dust-mass 
  ejections with the far-infrared (IR) and submillimeter observations with the 
  \textit{Herschel Space Observatory} has revealed that the dust content of AGB stars falls short 
  by an order of magnitude in explaining the observed dust content of the
  Magellanic Clouds. Hence, high-mass stars are likely a key source of 
  interstellar dust in these galaxies,
  and it is very important for us to understand their contribution 
  to the total dust-mass budget of the ISM of these galaxies.
  

  To quantify the dust mass and the dust composition, 
  we have choose to investigate the spectrum and  spectral energy distribution (SED)
  of an LBV, R71 (HDE269006) in the LMC as a case study.
  We chose LMC because of its proximity (50 kpc; Scaefer \cite{Schaefer-2008}).
  In addition, studies of this kind 
  are very difficult for the MWG because of confusion 
  along the line-of-sight and distance ambiguity. 
  One of the main motivations of this work is to complete the inventory of dust producers in the LMC.  
  The AGB and RSG stars have been analyzed by various authors 
  (e.g., Sargent et al. \cite{Sargent-et-al-2010}, \cite{Sargent-et-al-2011}; Riebel et al. \cite{Riebel-et-al-2012}), 
  but not the other massive stars.  
  This article is the first attempt to thoroughly analyze of this prime target using all current data,
  including the new photometric fluxes from the \textit{Herschel Space Telescope}.
  
  
  R71 (M$_{ZAMS}$ $\simeq$ 40 M$_{\odot}$; Lennon et al. \cite{Lennon-et-al-1993}) 
  is a very interesting LBV to study because 
  it has been observed by various telescopes in a wide range of wavelengths, which provided  
  spectroscopic data as well as photometric images.
  Various authors have studied either the spectral features 
  (e.g., Voors et al. \cite{Voors-et-al-1999}; Morris et al. \cite{Morris-et-al-2008};
  vanLoon et al. \cite{vanLoon-et-al-2010}; Waters \cite{Waters-2010})
  or photometric images (e.g., Bonanos et al. \cite{Bonanos-et-al-2009};
  Boyer et al. \cite{Boyer-et-al-2010}) of R71 separately
  and determined dust properties for different wavelength regions. We 
  have combined for the first time all the available photometric and spectroscopic data 
  of R71 for a wide wavelength range (0.36 -- 250\,$\mu$m). This gives us an excellent
  opportunity to analyze the data to accurately determine the total dust 
  mass and composition. 
  
  Voors et al. (\cite{Voors-et-al-1999}) have used the 1D radiative transfer model MODUST to 
  analyze the \textit{ISO} CAM and SWS spectrum of R71. Since the \textit{ISO} SWS spectrum between 
  30 -- 45\,$\mu$m was very noisy, their analysis was limited to the 6 -- 28\,$\mu$m  wavelength range. 
  These authors have identified the spectral PAH features at 6.2 and 7.7\,$\mu$m and amorphous silicate feature at
  10.0\,$\mu$m and a crystalline silicate feature at 23.5\,$\mu$m in the dust shell of R71. 
  From the mid-IR spectrum of R71 they estimate a dust mass of 0.02 M$_{\odot}$ and a time-averaged dust
  mass-loss rate of 7$\times$10$^{-6}$ M$_\odot$\,yr$^{-1}$ over $\approx$ 3000 yr.  
  But their model fails to reproduce the observation between 5 -- 15\,$\mu$m and also the 
  60 and 100\,$\mu$m \textit{IRAS} fluxes. By extrapolating the \textit{ISO} SWS data, these authors speculated 
  that R71 may have a second dust shell.
  
  We have chosen to re-investigate the dust composition and dust mass of R71 because of the availability of 
  \textit{Spitzer} IRS spectroscopic data and the \textit{Herschel} new photometric fluxes. Compared with 
  Voors et al. (\cite{Voors-et-al-1999}), we are able to combine all the available photometric 
  and spectroscopic data 
  of R71 over a wide wavelength range (0.36 -- 250\,$\mu$m; see Sect. 2). The better measurements 
  of the far-IR photometric fluxes are important 
  for an accurate dust-mass estimation. Combining the new data points provides a well-constrained model. 
  The \textit{Spitzer} IRS spectroscopic data of R71 have a much better S/N than the \textit{ISO} SWS data that were  
  used by Voors et al. (\cite{Voors-et-al-1999}). Because of the excellent  data quality, we have identified
  three new crystalline silicate (forsterite) features at 18.9, 27.5, and 33.6\,$\mu$m,  in addition to the 23.5\,$\mu$m
  feature that was identified by Voors et al. (\cite{Voors-et-al-1999}). 
  Our identification of four  crystalline spectral features
  firmly establishes the presence and importance of forsterite in the dust shell of R71 (see Sect. 3). 
  Compared with Voors et al. (\cite{Voors-et-al-1999}),
  we have used the more recent and sophisticated 3D radiative transfer model MCMax (Min et al. \cite{Min-et-al-2009})
  for our analysis. We are able to fit the observational data in general very well compared with 
  Voors et al. (\cite{Voors-et-al-1999}).
   Additionally, our analysis allows us to derive a first estimate for the overall contribution 
  from the LBVs to the total dust-mass budget of the ISM of the LMC. 
  
  The organization of this paper is as follows:
  all the
  observational (photometric and spectroscopic) data are summarized 
  in Sect. 2. In Sect. 3, we discuss the 
  spectral dust features and possible potential mineral carriers of those features.
  We fit the entire SED of R71 using the 3D radiative transfer model MCMax. 
  The results from our model calculation are shown
  in Sect. 4. Finally, Sect. 5 contains the discussion of our results, 
  and in Sect. 6 we conclude our findings.
  

 \section{Observations}
    \label{Obs}

\begin{figure}
   \centering
   \includegraphics[width=\hsize]{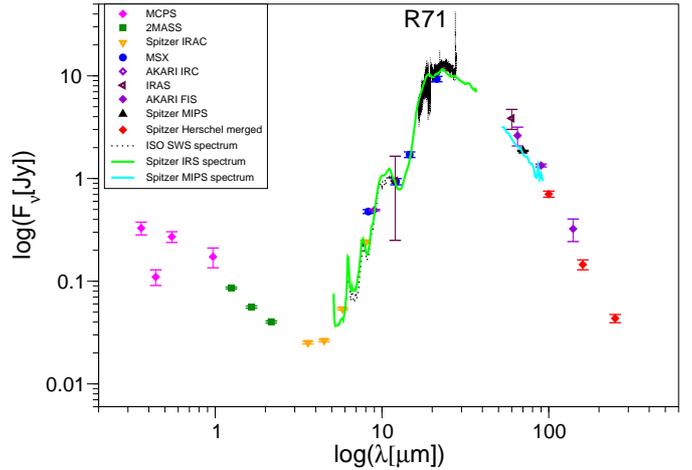}
      \caption{All observation data of R71 obtained from various telescopes. 
               }
         \label{R71_obs_error}
   \end{figure}

  We obtained optical to far-IR photometric fluxes and spectroscopic data 
  of R71 from various telescopes. All the observational data we used for 
  our analysis are shown in Fig.~\ref{R71_obs_error}.
  Optical U (0.36\,$\mu$m), B (0.44\,$\mu$m),
  V (0.55\,$\mu$m), and I (0.97\,$\mu$m) band photometric fluxes
  were obtained from the \textit{Magellanic Clouds Photometric Survey}
  (MCPS; Zaritsky et al. \cite{Zaritsky-et-al-1997}) and J (1.25\,$\mu$m), 
  H (1.65\,$\mu$m), K (2.17 \,$\mu$m) fluxes from the \textit{Two Micron All-Sky Survey}
  (2MASS; Skrutskie et al. \cite{Skrutskie-et-al-2006}) through the
  \textit{Surveying The Agents of Galaxy Evolution Spitzer Legacy program}
  (SAGE; Meixner et al. \cite{Meixner-et-al-2006}) catalog. Similarly, we  
  obtained \textit{Spitzer Space Telescope} (Werner et al. \cite{Werner-et-al-2004}) 
  InfraRed Array Camera (IRAC; Fazio et al. \cite{Fazio-et-al-2004}) photometric 
  fluxes at 3.6, 4.5, 5.8, and 8.0\,$\mu$m from the SAGE catalog.
  
  The mid-IR photometric fluxes at A (8.28\,$\mu$m), C (12.13\,$\mu$m),
  D (14.65\,$\mu$m), E (21.3\,$\mu$m) band were obtained from the
  \textit{Midcourse Space Experiment} (MSX; Egan et al. \cite{Egan-et-al-2001}) 
  survey images and source catalogs. For completeness, we included 
  the mid-to far-IR photometric fluxes at 12.0, 25.0, and 
  60.0\,$\mu$m from  the 
  \textit{InfraRed Astronomical Satelite} (IRAS; Neugebauer et al. \cite{Neugebauer-et-al-1984}) 
  from the IRAS catalog (Beichman et al. \cite{Beichman-et-al-1988}).
  
  We also obtained mid-to far-IR photometric fluxes 
  from the Infrared Astronomical Satellite \textit{AKARI} 
  (Murakami et al. \cite{Murakami-et-al-2007}) from the Infrared Camera (IRC) 
  point source catalog (Kataza et al. \cite{Kataza-et-al-2010}) at 9.0 and 18.0\,$\mu$m 
  and the \textit{Far-Infrared Surveyor} (FIS) Bright Source catalog 
  (Yamamura \cite{Yamamura-2010}) at 65.0, 90.0, and 140.0\,$\mu$m. 
  The far-IR photometric fluxes were obtained from the
  \textit{Spitzer Space Telescope} Multiband Imaging Photometer (MIPS; Rieke et al. \cite{Rieke-et-al-2004}) 
  at 70\,$\mu$m (at 24\,$\mu$m, the flux is saturated) from the SAGE catalog 
  (Meixner et al. \cite{Meixner-et-al-2006}). 
  Recently, the photometric fluxes from the \textit{Spitzer} MIPS and \textit{Herschel}
  Photodetector Array Camera and Spectrometer (PACS; Poglitsch et al. 
  \cite{Poglitsch-et-al-2010}) and from the Spectral and Photometric Imaging REceiver 
  (SPIRE; Griffin et al. \cite{Griffin-et-al-2010}) have been merged 
  to improve both the quality of the data and the extraction process
  (e.g., Meixner et al. \cite{Meixner-et-al-2013}; Seale et al. \cite{Seale-et-al-2014}).
  We obtained these new photometric merged fluxes
  at 100.0, 160.0, and 250.0\,$\mu$m of R71. The fluxes at 350.0 and 500.0\,$\mu$m are not
  detected by them.
  These new photometric fluxes better reflect both 
  the flux and flux uncertainty of R71 compared with the  \textit{Herschel} PACS and SPIRE data,
  that were used by Boyer et al. (\cite{Boyer-et-al-2010}).

  To analyze the spectral dust features of R71,
  we also obtained the spectroscopic data
  in the region of 5.2 -- 38\,$\mu$m from the \textit{Spitzer} InfraRed Spectrograph 
  (IRS; Houck et al. \cite{Houck-et-al-2004}) and the \textit{Spitzer} MIPS spectrum 
  in the region of 52 -- 97\,$\mu$m (SAGESpec; Kemper et al. \cite{Kemper-et-al-2010}; 
  van Loon et al. \cite{vanLoon-et-al-2010}). 
  We also included mid-IR spectra
  from the Infrared Space Observatory (\textit{ISO}; Kessler et al. \cite{Kessler-et-al-1996}) 
  Imaging Photo-Polarimeter (ISOPHOT; Lemke et al. \cite{Lemke-et-al-1996}) 
  for the region between 6 -- 16\,$\mu$m and from the Short Wavelength Spectrometer 
  (SWS; de Graauw et al. \cite{deGraauw-et-al-1996}) for the region between 16 -- 28\,$\mu$m. 
  The \textit{ISO} data match the \textit{Spitzer} data very well.
   The values of all the photometric fluxes ($F_\nu$) 
  at different wavelengths from various instruments are listed in Table~\ref{R71_obs},
  along with $\sigma$ values and corresponding references.
  
%
  

 \section{Guided tour through the spectrum of R71}
    \label{SED}  

   In this section, we mainly focus on the IR spectroscopic data of R71
   to identify the spectral dust features and the potential 
   dust mineral carrier of these features.

   \begin{figure}
   \centering
   \includegraphics[width=\hsize]{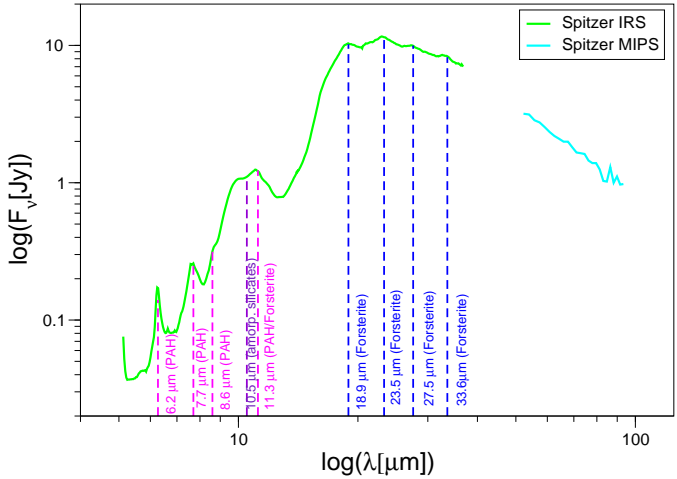}
      \caption{\textit{Spitzer} IRS and MIPS spectra of R71. 
               We plot the nine spectral dust features (indicated by dashed straight lines), 
               along with the potential dust minerals as carriers of these features.}
         \label{R71_spectroscopy}
   \end{figure}
   
   In Fig.~\ref{R71_spectroscopy}, we present only the spectroscopic data of R71 
   obtained from \textit{Spitzer} IRS and MIPS instruments (for references see Table~\ref{R71_obs}).
   Because the \textit{ISO} SWS spectrum matches the \textit{Spitzer} IRS
   spectrum very well (see Sect. 2; Fig.~\ref{R71_obs_error}), we did not include the \textit{ISO} 
   spectrum in Fig.~\ref{R71_spectroscopy}.
   
   We note nine spectral dust features in total  at 6.2, 7.7, 8.6, 10.5, 11.3,
   18.9, 23.5, 27.5, and 33.6\,$\mu$m in the dust shell of R71. 
   Evolved high-mass stars (with the exception of WR stars) are generally dominated by an
   oxygen-rich chemistry ($C/O < 1$), and therefore the condensation of carbon-rich ($C/O > 1$) 
   dust species is not very obvious. R71 shows
   both silicates and carbonaceous material in its ejecta.
   
   
   The \textit{Spitzer} IRS spectrum of R71 shows clear evidence of spectral features 
   at 6.2, 7.7, 8.6 (possibly), and 11.3\,$\mu$m. These features were also identified by 
   Voors et al. (\cite{Voors-et-al-1999}) in the \textit{ISO} SWS spectrum of R71
   (also shown in Fig.~\ref{R71_obs_error}, black dotted line), although the 3.3\,$\mu$m 
   PAH emission feature is not detected in the \textit{ISO} SWS spectrum.
   These features are attributed to C-H and C-C stretching and bending modes in large 
  ($\approx$ 50 C-atom) polycyclic aromatic hydrocarbon (PAH) molecules 
  (e.g., Allamandola et al. \cite{Allamandola-et-al-1985}; 
  Leger \& Puget \cite{Leger-Puget-1985}; Tielens \cite{Tielens-2008}). 
  
  We note that the 
  11.3\,$\mu$m feature can also have a contribution from 
  crystalline silicate (forsterite: Mg$_2$SiO$_4$; Molster et al. \cite{Molster-et-al-2002a};  
  Guha Niyogi et al. \cite{GuhaNiyogi-et-al-2011}). The PAH features
  indicate C-rich dust.
  
\begin{figure}
   \centering
    \includegraphics[width=\hsize]{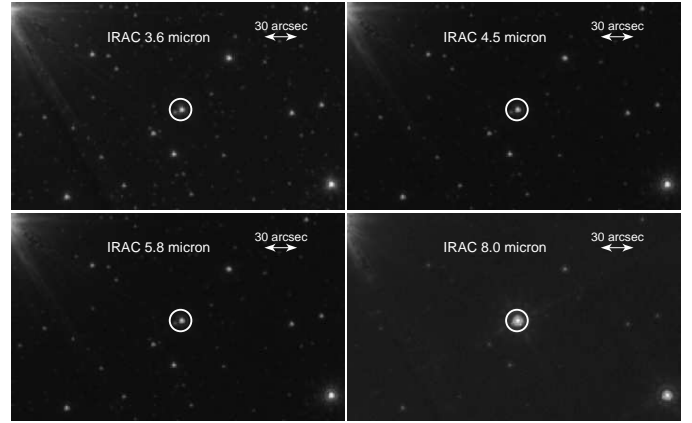}
       \caption{\textit{Spitzer} IRAC images at 3.6, 4.5, 5.8, and 8.0\,$\mu$m of R71, which clearly show that 
               the PAH features are associated with R71. R71 is a point source taken from the \textit{Spitzer} SAGE catalog
               and is located inside the white
               circle; the radius of the circle is 10 arcsec. The extent in 8.0\,$\mu$m image reflects the PSF size.}
        \label{R71_IRAC_images}
   \end{figure}

  To clarify whether these PAH features are associated with
  R71 or emission from the background ISM, we obtained the photometric 
  images from \textit{Spitzer} IRAC at 3.6, 4.5, 5.8, and 8.0\,$\mu$m.
  The images of R71 are shown in Fig.~\ref{R71_IRAC_images}. R71 is a
  point source, and the extent shown in Fig.~\ref{R71_IRAC_images} reflects
  the point spread function (PSF) size. From IRAC 8.0\,$\mu$m image it is 
  very clear that the apparent shape is a typical bright \textit{Spitzer} IRAC
  PSF for the LMC (FWHM = 3.6 arcsec; Meixner et al. \cite{Meixner-et-al-2006}).
  From the photometric images, we are confident that these spectral features
  are associated with R71.

  The \textit{Spitzer} IRS spectrum reveals a broad feature 
  between 8 -- 12\,$\mu$m. Within this broad feature, we identify 
  a prominent amorphous silicate feature
  at 10.5\,$\mu$m and  a crystalline silicate/PAH feature at 
  11.3\,$\mu$m (as already discussed above). 
  This silicate feature
  indicates O-rich dust. This silicate feature was 
  previously detected in many astrophysical environments, including the solar system 
  and extra-solar planetary 
  systems (Mann et al. \cite{Mann-et-al-2006}), 
  the circumstellar regions  of both young stellar objects, 
  AGB stars, and planetary nebulae
  (e.g., Speck et al. \cite{Speck-et-al-2000}; Casassus et al. \cite{Casassus-et-al-2001});
   many lines of sight through the interstellar medium in our own galaxy 
   (Chiar et al. \cite{Chiar-et-al-2007}), and
   in nearby and distant galaxies (Hao et al. \cite{Hao-et-al-2005}).
   
\begin{figure}
   \centering
   \includegraphics[width=\hsize]{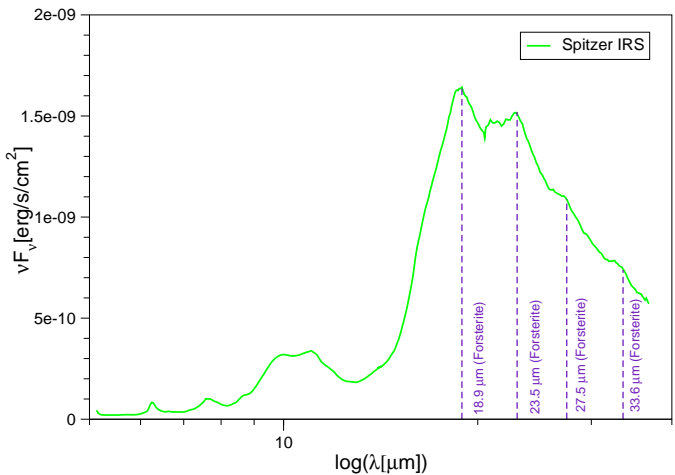}
      \caption{\textit{Spitzer} IRS spectrum of R71. 
               These are the same data as Fig.~\ref{R71_spectroscopy}, but here the
               y-axis is plotted in linear scale to
               emphasize the crystalline silicate (forsterite) features at 18.9, 23.5, 27.5, and 33.6 $\mu$m.}
         \label{R71_spectroscopy_nuFnu}
   \end{figure}

   
   The spectrum rises sharply at 12\,$\mu$m and has a broad peak between 12 -- 37\,$\mu$m. 
   Since the \textit{ISO} SWS spectrum between 30 -- 45$\mu$m was very noisy, 
   Voors et al. (\cite{Voors-et-al-1999}) only identified one crystalline silicate feature at
   23.5\,$\mu$m. Because of the availability of high-quality \textit{Spitzer} IRS spectrum of R71, 
   we identified four crystalline silicate spectral features at 18.9, 23.5, 27.5, and 33.6\,$\mu$m in the dust shell of R71 
   (see Fig.~\ref{R71_spectroscopy} and Fig.~\ref{R71_spectroscopy_nuFnu}).
   These features are attributed to 
   crystalline silicates: forsterite (Mg$_2$SiO$_4$).
   These features have also been identified 
   before in various astrophysical objects (e.g., young stellar objects, AGB stars, post-AGB stars, 
   planetary nebulae, high-mass stars) by various authors (e.g., Waters et al. 
   \cite{Waters-et-al-1996}; Waelkens et al. \cite{Waelkens-et-al-1996};
   Molster et al. \cite{Molster-et-al-2002a};
   Guha Niyogi et al. \cite{GuhaNiyogi-et-al-2011}). The steep rise between 
   16 and 19.5\,$\mu$m stems from
   the contribution of both the amorphous and crystalline silicates.
   
   That there are PAHs and silicates in the 
   dust shell of R71 is no unique finding. 
   The presence of PAHs in an O-rich environment have also been detected 
   in the \textit{ISO} SWS spectra of other Galactic LBVs,  
   for example, AG Car (Voors et al. \cite{Voors-et-al-2000}) and 
   HD 168625 (Skinner \cite{Skinner-1997}), and in the planetary nebula (PN) NGC 6302
   (Beintema \cite{Beintema-1998}). Recently, mixed chemistry 
   has also been detected in PNe toward the Galactic bulge by
   analyzing \textit{Spitzer} IRS spectra 
   (e.g., Gutenkunst et al. \cite{Gutenkunst-et-al-2008}; 
   Guzman-Ramirez et al. \cite{Guzman-Ramirez-et-al-2011}).
   
   The origin of PAH emission in the dust shell of R71 is  unclear.
   The presence of both the silicates and carbonaceous compounds suggests
   that there were perhaps two almost simultaneous mass-loss events 
   associated with the shell: 
   one that led to silicate condensation, and another that led to carbon dust. 
   This is a common phenomenon for LBVs, 
   but it does not necessarily imply O-rich and C-rich ejecta. 
   Some O-rich ejecta also show the two dust types in their ejecta (e.g., Sylvester \cite{Sylvester-1999}).
   Alternatively, PAH formation reflects the condensation of carbonaceous 
   compounds on iron-rich grains through Fischer-Tropsch-like reactions 
   (e.g., Kress \& Tielens \cite{Kress-Tielens-1996}) 
   and the subsequent release of these species into the gas phase.


  \section{Dust model for R71}
     \label{model}

   After identifying the spectral dust features and the dust composition
   in the dust shell of R71, the next step is to model the SED of R71.
   
   We computed the model spectrum of R71 using the Monte Carlo radiative
   transfer code MCMax (Min et al. \cite{Min-et-al-2009}).
   This code is based on a 3D Monte Carlo 
   radiative transfer model (e.g., Bjorkman \& Wood \cite{Bjorkman-Wood-2001};
   Niccolini et al. \cite{Niccolini-et-al-2003}; Pinte et al. \cite{Pinte-et-al-2006}).
   In this method photon packages emerging from the central 
   star are traced through the dust shell, allowing them to undergo
   scattering, absorption, and re-emission events caused by the dust they encounter on their path.
   For a detailed description of the MCMax code, we refer to Min et al. (\cite{Min-et-al-2009}).
   Our choice of using this code over others (e.g., DUSTY, 2-DUST) was made simply because 
   it is accurate at all optical depths and 
   sets virtually no restrictions on the spatial 
   distribution of matter or on the optical properties of the dust grains. Last not least, 
   it is a self-consistent method. 
   
   \subsection{Stellar parameters}
     \label{stellar}
   
   For our model calculation, we treated the LBV as a central star 
   enclosed by a spherically symmetric dust shell. The central star was  
   treated as a black body defined by its radius and temperature.
   We adopted a stellar temperature ($T_{eff}$) = 17,000\,K and luminosity
   ($L_\star$) = $7\times10^5$ L$_\odot$, stellar radius ($R_\star$) = 95 R$_\odot$, 
   and stellar mass ($M_\star$) = 20 M$_\odot$. The values of these stellar parameters
   were taken from Lennon et al. (\cite{Lennon-et-al-1993}) and are listed 
   in Table~\ref{R71_model}. 
   
   \subsection{Dust-shell parameters}
     \label{shell}
     
   We assumed that the density distribution inside the dust shell is spherically
   symmetric and time independent with an $r^{-2}$ radial dependence. We adopted a  
   dust-to-gas ratio ($f_{d/g}$) of 1/500 for the LMC, as applied by other authors
   (e.g., van Loon et al. \cite{vanLoon-et-al-1999}; Sargent et al. \cite{Sargent-et-al-2011})
   to calculate the mass-loss late of RSGs in the LMC. This value of $f_{d/g}$ is 
   believed to be representative of the moderately low LMC metallicity.
   
    \begin{figure}
    \centering
    \includegraphics[width=\hsize]{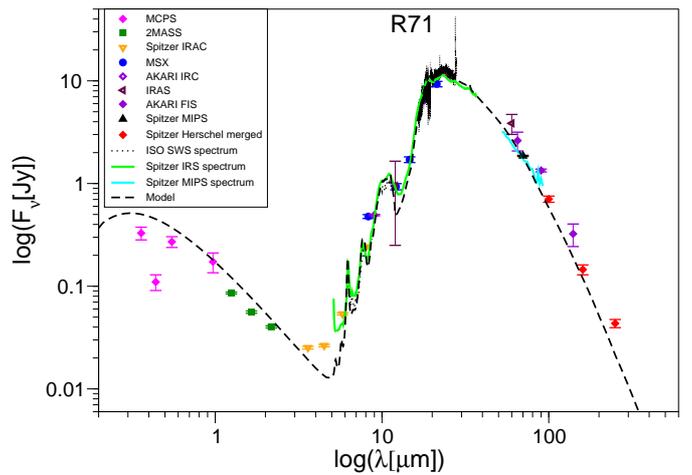}
      \caption{Best-fit model of R71 along with all the observational data. The model fits very 
               well over the entire wavelength region. 
               Our model is shown as a black dashed curve.
               All the input and fitted parameters are listed in Table~\ref{R71_model}.}
         \label{R71_obs_model}
   \end{figure}
   
   The best-fit model along with all the observational data is shown in Fig.~\ref{R71_obs_model}
   for the entire wavelength region. The model fits the 
   SED in general very well. From our best-fit models we derive that
   it requires a dust shell of inner radius ($R_{in}$) = $1.0\times10^4$ A.U., 
   outer radius ($R_{out}$) = $6.0\times10^4$ A.U., and a total dust mass
   ($M_{dust}$) = 0.01 M$_\odot$. These dust shell parameters are also listed in Table~\ref{R71_model}. 
   From these results, we interpret that R71 has an optically thin dust shell. We note that 
   the dust mass might be underestimated because there are dense, optically thick dust clumps.
   The far-IR fluxes will probably detect most of the dust and are not expected to vary significantly 
   from our derived dust mass.

    \subsection{Dust composition}
     \label{composition}
     
\begin{figure}
    \centering
   \includegraphics[width=\hsize]{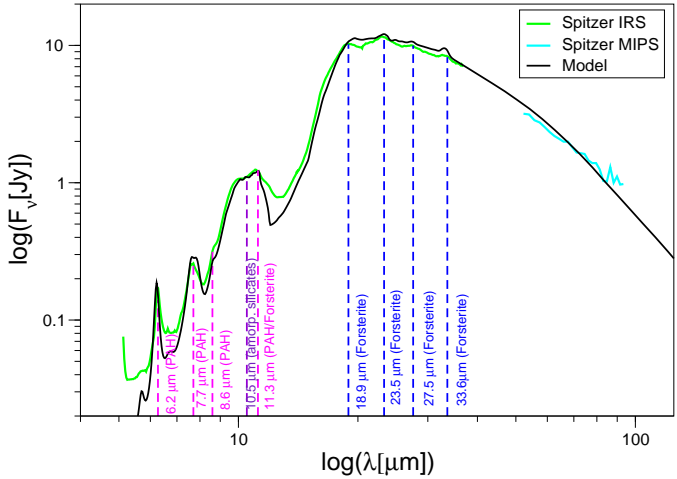}
      \caption{Best-fit dust model plotted against the spectroscopic observational data of R71, 
               obtained by \textit{Spitzer} IRS and MIPS. 
               In general, the model fits the PAHs and silicate features well.}
         \label{R71_spectroscopy_model}
   \end{figure}

   In Sect. 3, we have identified the spectral dust features in the dust shell of R71
   and the potential mineral carriers for these features. For the dust composition, 
   we used mixture of amorphous and crystalline silicates and PAHs. For the amorphous dust we 
   used two types of amorphous silicates (MgSiO$_3$ and MgFeSiO$_3$), for crystalline
   silicate we used forsterite (Mg$_2$SiO$_4$). As a smooth 
   continuum opacity we also included metallic iron (Kemper et al. \cite{Kemper-et-al-2002}) 
   in our model calculation. We used the optical constants 
   ($n, k$; for reference see Table~\ref{R71_model}) 
   of these minerals to calculate their absorption cross-section (C$_{abs}$).
   In Fig.~\ref{R71_spectroscopy_model} we show 
   the spectroscopic data of R71 by \textit{Spitzer} IRS and MIPS with our best-fit model.
   
   From our best-fit model we derived that 
   the dust is mainly dominated by amorphous silicates: MgSiO$_3$ (63\%), and 
   MgFeSiO$_3$ (30\%), and with some crystalline silicate: Mg$_2$SiO$_4$ (3\%), metallic iron (3\%),
   and a very tiny amount of PAHs (0.1\%). The dust composition with their mass-abundances that we 
   used in our model are listed in Table~\ref{R71_model}.
   From the SED, we calculated a ratio of the PAH emission to the total = 0.02, 
   which in turns leads to a mass abundance ($m_{PAH}/m_{Dust}$) of $\approx 0.05$ (Tielens \cite{Tielens-2005}).
   Depending on the formation mechanism of the PAHs in the O-rich environment, it is quite 
   plausible that they have a much lower abundance than the bulk of the dust and the gas density.

   From Fig.~\ref{R71_spectroscopy_model}, we can see that in general our model fits the PAHs 
   and silicate features much better than the model fit of Voors et al. (\cite{Voors-et-al-1999}). 
   The only discrepancy of the model with the observation is at around 7.0 and 12.0\,$\mu$m. We suggest that the amorphous 
   silicate profile (which was calculated from laboratory-measured optical constants; see Table~\ref{R71_model}) 
   may not provide an entirely suitable representation of the actual interstellar and/or circumstellar silicate profile. 
   In fact, the profile of interstellar/circumstellar silicates is only poorly determined in the 12.0\,$\mu$m region 
   (Chiar \& Tielens \cite{Chiar-Tielens-2006}).
   We also note that Voors et al. (\cite{Voors-et-al-1999}) were unable to model the 12 -- 15\,$\mu$m 
   region and attributed this to a 
   missing warm component in their model calculation. The LBV HD168625 provides another example of a source with the 
   same problem. It has been speculated that in this source the problem reflects transiently heated very small 
   grains (O'Hara et al. \cite{O'Hara-et-al-2003}). We note that our overall good fit of the R71 spectrum argues against 
   including other dust materials because they would contribute to the emission over a much broader wavelength region than the 
   quite limited wavelength ranges where our model fails to provide a satisfactory fit.

   \subsection{Dust-grain size distribution}
     \label{size}
   
   For both amorphous and crystalline silicates and metallic iron we used 
   grain sizes between 0.01 -- 0.1\,$\mu$m, of grain size distribution ($n(a) \propto  a^{-q}$, 
   where $q = 3.5$) adopted from Mathis, Rumple \& Nordsieck 
   (\cite{Mathis-Rumple-Nordsieck-1977}).
   The very low PAH emission in the dust shell of R71 indicates a population of
   very small grains that are probably not in thermal equilibrium. For PAHs, we used
   dust grain sizes of 0.001 -- 0.0011\,$\mu$m (i.e., 540 C-atoms). 
   We tested models with grain-size distributions 
   of 0.01 -- 1.0\,$\mu$m or 0.01 -- 10.0\,$\mu$m for silicate particles 
   that required significantly different 
   dust-masses of the shell, but still did not fit as well, as shown in Fig.~\ref{R71_obs_model}.
   These dust size distributions are also listed in Table~\ref{R71_model}.

   \section{Discussion}
     \label{discussion}
     
     Our derived $M_{dust}$ = 0.01 M$_\odot$ of R71 (see Sect. 4.2) is an order of 
     magnitude lower than previously derived by Boyer et al. 
     (\cite{Boyer-et-al-2010}) using \textit{Herschel} PACS and SPIRE data. 
     This is because these authors had taken into account the submm 
     emission (at 350, 500\,$\mu$m) in their analysis, but these photometric 
     fluxes were later discarded because they were very weak 
     (as discussed in Sect. 2; Meixner et al. \cite{Meixner-et-al-2013}; 
     Seale et al. \cite{Seale-et-al-2014}).
     Our derived $M_{dust}$ for R71 has a similar order of magnitude as the 
     previously derived $M_{dust}$ of R71 by Voors et al. (\cite{Voors-et-al-1999}),
     and also WRA 751 in MWG (Voors et al. \cite{Voors-et-al-2000}) 
     obtained from an analysis of the \textit{ISO} SWS spectrum.
     In the MWG, the two well-studied LBVs are AG Car and 
     $\eta$ Car. The dust mass injected by AG Car is $\simeq0.1$ M$_{\odot}$ 
     (Voors et al. \cite{Voors-et-al-2000}) and that of $\eta$ Car is  $\simeq0.4$ M$_{\odot}$ 
     (Gomez et al. \cite{Gomez-et-al-2010}),
     which is an order of magnitude higher than the dust mass ejected by R71.
     
     From the derived $M_{dust}$ = 0.01 M$_\odot$
     and using $f_{d/g}$ = 1/500 for the LMC (see Sect. 4.2),
     we derived a total ejected mass ($M_{total}$) of $\approx$ 5 M$_\odot$.
     This result agrees with the empirical ``laws''  
     by Stothers \& Chin (\cite{Stothers-Chin-1995}).
     These authors have shown that the LBVs with luminosity log (L/L$_\odot$) $\simeq$ 5.7 
     (for R71, log (L/L$_\odot$)= 5.85), have episodes of sudden mass loss of
     4 $\pm$ 1 M$_\odot$. Additionally, with the availability of 
     far-IR photometric fluxes (see Fig.~\ref{R71_obs_error}), 
     we established that there is no second dust shell with a dust mass 
     of 0.3 M$_{\odot}$, as speculated by Voors et al. (\cite{Voors-et-al-1999}).

     The optical spectroscopic observational data  of R71 revealed  
     that the $v_{exp}$ of its circumstellar matter shows a complex 
     behavior. From the literature, we found that the value of $v_{exp}$
     varies between $\approx$ 10 -- 130 km/s (e.g., Wolf et al. \cite{Wolf-et-al-1981};
     Stahl \& Wolf \cite{Stahl-Wolf-1986}; Mehner et al. \cite{Mehner-et-al-2013}).
     Here, we adopted the average value of $\approx 60$ km/s
     (Wolf et al. \cite{Wolf-et-al-1981}).
     From our derived
     dust-shell parameters ($R_{in}, R_{out}, M_{dust}$; see Sect. 4.2)
     and using the average expansion velocity of 60 km/s, we derived the time-averaged 
     dust mass-loss rate of 2.5$\times$10$^{-6}$ M$_\odot$\,yr$^{-1}$ over $\approx$ 4000 years.
     This result falls into the range of 
     $10^{-7} < \dot{M} < 10^{-5}$\,M$_\odot$\,yr$^{-1}$ (at its minimum
     and maximum phase), as predicted by Wolf et al. (\cite{Wolf-et-al-1981}).
     Our derived mass-loss rate for R71 is on the same order of magnitude as
     derived for R71 and WRA 751 by Voors et al. (\cite{Voors-et-al-1999}, \cite{Voors-et-al-2000}). 
     $\eta$ Car is well-known for its dramatic outbursts; 
     its most famous mass ejection occurred in the early 19th century and is known 
     as homunculus. Gomez et al. (\cite{Gomez-et-al-2010}) estimated 
     the dust mass in the stellar wind to be 0.4 $\pm$ 0.1 M$_\odot$, which translates
     into a mass loss greater than 40$_\odot$ over the past 1000 years.
     It is clear, however that not all LBVs have dust mass-loss rates as high as $\eta$ Car,
     and fewer than half of the objects have notable excesses
     (van Genderen \cite{van-Genderen-2001}).
     
     There are six confirmed dusty LBVs in the LMC, including R71 (Humphreys \& Davidson 
     \cite{Humphreys-Davidson-1994}). Another five or six have been suggested as candidates in the 
     literature (e.g., Weis \cite{Weis-2003}). To estimate 
     the overall contribution from the LBVs to the total dust-mass budget 
     of the ISM of the LMC, we assumed that their time-averaged 
     dust mass-loss rate is approximately similar ($\approx$ 2.5$\times$10$^{-6}$ M$_\odot$\,yr$^{-1}$).
     The total mass-loss rate for six LBVs (including R71) is then $\approx$ 
     1.3$\times$10$^{-5}$ M$_\odot$\,yr$^{-1}$. For the LMC, this is comparable to the mass-loss rate
     by all the AGBs and RSGs ($\approx$ 10$^{-5}$ M$_\odot$\,yr$^{-1}$;
     Srinivasan et al. \cite{Srinivasan-et-al-2009}; Matsuura et al. \cite{Matsuura-et-al-2009};
     Tielens \cite{Tielens-2014}). This rough estimation implies that LBVs as a class may be an important
     source of interstellar dust.
     Hence, a more detailed investigation of the total dust production by this class of stars 
     and determination of their contribution to the dust budget of the LMC is a promising task.
      


  \section{Conclusions}
    \label{Con}

  We have analyzed the IR spectrum and SED of the LBV R71 in the LMC.
  
  We conclude that:
  \begin{enumerate}
  \item{
  The SED of R71 shows that the short-wavelength region is dominated by the stellar emission,
  the long-wavelength region by thermal emission of dust.}
  
  \item{
  R71 is enclosed by an optically thin dust shell. We adopted a
  density distribution that is spherically
  symmetric and time independent with an $r^{-2}$ radial dependence.}
  
  \item{
  We note nine spectral dust features in total at 6.2, 7.7, 8.6, 10.5, 11.3, 18.9, 23.5, 27.5, and
  33.6\,$\mu$m from \textit{Spitzer} IRS and MIPS spectra. Among them, we identified three new 
  crystalline silicate features at 18.9, 27.5, and 33.6\,$\mu$m.}
  
  \item{
  The chemistry of the dust shell is mainly dominated by amorphous silicates, 
  with some crystalline silicates, metallic iron, and with a very tiny amount of PAHs.}
  
  \item{
  R71 shows both silicates and carbonaceous material in its ejecta.
  Possibly, two almost simultaneous mass-loss events 
  associated with the shell created separate silicate and carbon dust condensations.}
  
  \item{
  We used the 3D radiative transfer code MCMax to compute the model spectrum. The model fits 
  the overall SED and spectral features of R71 very well. 
  From our model calculation, we derived dust shell parameters,
  dust composition, and dust mass.}
  
  \item{
  We derived $M_{total}$ $\approx$ 5.0 M$_\odot$ from the observed $M_{dust}$ = 0.01 M$_\odot$.
  This result agrees with predictions for similar types of LBVs as R71.}
  
  \item{
  From the dust-shell parameters, we derived a time-averaged 
  dust mass-loss rate of 2.5$\times$10$^{-6}$ M$_\odot$\,yr$^{-1}$. 
  The mass-loss rate of R71 falls into the typical 
  range of mass-loss rates of LBVs.}
  
  \item{
  From the derived mass-loss rate of R71, we extrapolated a mass-loss rate of 
  1.3$\times$10$^{-5}$ M$_\odot$\,yr$^{-1}$ for six confirmed dusty LBVs (including R71) in the LMC,
  which is comparable to the mass-loss rate of all the AGBs and RSGs.}
  
  \item{
  A more detailed investigation of 
  the total dust production by this class of stars and a determination of their contribution to the dust 
  budget of the LMC is a promising task.}

  \end{enumerate}   

\begin{acknowledgements}
      We are grateful to the referee, whose comments improved the quality of the article.
      We also thank  Catharinus Dijkstra and Martha Boyer for their fruitful 
      comments and feedback. 
      Studies of interstellar dust and interstellar chemistry at Leiden 
      Observatory are supported through advanced-ERC grant 246976 from the European Research 
      Council, through a grant by the Dutch Science Agency, NWO, as part of the Dutch 
      Astrochemistry Network, and through the Spinoza premie from the Dutch Science Agency, NWO.
      
\end{acknowledgements}


%
 \clearpage
 \begin{table}[t]
   \centering
 \caption{Observed photometric and spectroscopic fluxes at various wavelengths from various instruments.}    
 \small
 \label{R71_obs}     
 \begin{tabular}{l c c c c}  
 \hline\hline                 
 Telescopes & $\lambda$ (\,$\mu$m) & $F_\nu$ (Jy) & $\sigma$ & References\\    
 \hline                        
   MCPS & 0.36 & 0.33 & 0.047 & Zaritsky et al. (\cite{Zaritsky-et-al-1997})\\
   MCPS & 0.44 & 0.11 & 0.019 & Zaritsky et al. (\cite{Zaritsky-et-al-1997})\\
   MCPS & 0.55 & 0.27 & 0.032 & Zaritsky et al. (\cite{Zaritsky-et-al-1997})\\	
   MCPS & 0.97 & 0.17 & 0.038 & Zaritsky et al. (\cite{Zaritsky-et-al-1997})\\
   2MASS& 1.25 & 0.08 & 0.002 & Skrutskie et al. (\cite{Skrutskie-et-al-2006})\\
   2MASS& 1.65 & 0.06 & 0.001 & Skrutskie et al. (\cite{Skrutskie-et-al-2006})\\		
   2MASS& 2.17 & 0.04 & 0.001 & Skrutskie et al. (\cite{Skrutskie-et-al-2006})\\  
   Spitzer IRAC & 3.6 & 0.025 & 0.008 & Fazio et al. (\cite{Fazio-et-al-2004})\\
   Spitzer IRAC & 4.5 & 0.026 & 0.0007& Fazio et al. (\cite{Fazio-et-al-2004})\\
   Spitzer IRAC & 5.8 & 0.054 & 0.001 & Fazio et al. (\cite{Fazio-et-al-2004})\\
   Spitzer IRAC & 8.0 & 0.241 & 0.007 & Fazio et al. (\cite{Fazio-et-al-2004})\\
   MSX & 8.28 & 0.48 & 0.021 & Egan et al. (\cite{Egan-et-al-2001})\\
   AKARI IRC & 9.0 & 0.49 & 0.008 & Kataza et al. (\cite{Kataza-et-al-2010})\\
   IRAS & 12.0	& 0.95 & 0.7 & Beichman et al. (\cite{Beichman-et-al-1988})\\
   MSX & 12.13	& 0.93 & 0.064 & Egan et al. (\cite{Egan-et-al-2001})\\
   MSX & 14.65 & 1.71 & 0.108 & Egan et al. (\cite{Egan-et-al-2001})\\
   AKARI IRC & 18.0 & 6.47 & 0.088 & Kataza et al. (\cite{Kataza-et-al-2010})\\
   MSX & 21.3 &	9.31 & 0.559 & Egan et al. (\cite{Egan-et-al-2001})\\
   IRAS & 25.0 & 11.84 & 0.65 & Beichman et al. (\cite{Beichman-et-al-1988})\\
   IRAS & 60.0	& 3.86 & 0.85 & Beichman et al. (\cite{Beichman-et-al-1988})\\
   AKARI FIS & 65.0 & 2.62 & 0.544 & Yamamura (\cite{Yamamura-2010})\\
   Spitzer MIPS & 70.0 & 1.86 & 0.024 & Meixner et al. (\cite{Meixner-et-al-2006})\\
   AKARI FIS & 90.0 & 1.34 & 0.043 & Yamamura (\cite{Yamamura-2010})\\
   Spitzer Herschel merged & 100.0 & 0.71 & 0.049 & Meixner et al. (\cite{Meixner-et-al-2013})\\ 
   AKARI FIS & 140.0 & 0.32 & 0.08 & Yamamura (\cite{Yamamura-2010})\\ 
   Spitzer Herschel merged & 160.0 & 0.15 & 0.016 & Meixner et al. (\cite{Meixner-et-al-2013})\\ 	
   Spitzer Herschel merged & 250.0 & 0.04 & 0.004 & Meixner et al. (\cite{Meixner-et-al-2013})\\ 
   ISO PHT,SWS & 6.0 -- 28.0 & Spectroscopy & -- & Lemke et al. (\cite{Lemke-et-al-1996}); de Graauw et al. (\cite{deGraauw-et-al-1996})\\
   Spitzer IRS & 5.2--38.0 & Spectroscopy & -- & Houck et al. (\cite{Houck-et-al-2004})\\
   Spitzer MIPS & 52 -- 97 & Spectroscopy & -- & Kemper et al. (\cite{Kemper-et-al-2010}); van Loon et al. (\cite{vanLoon-et-al-2010})\\
 \hline
 \end{tabular}
 \end{table}

 \clearpage
 \begin{table}[t]
\caption{Input and fit parameters of R71.}             
\label{R71_model}      
\centering                          
\begin{tabular}{l c c c}        
\hline\hline                 
Input parameters\\    
\hline    
Stellar parameters & Symbol & Value & Reference\\
\hline
   Initial mass & $M_{ZAMS}$ & 40 M$_\odot$ & Lennon et al. (\cite{Lennon-et-al-1993})\\ 
   Mass & $M_\star$ & 20 M$_\odot$ & Lennon et al. (\cite{Lennon-et-al-1993})\\ 
   Radius & $R_\star$  & 95 R$_\odot$ & Lennon et al. (\cite{Lennon-et-al-1993})\\ 
   Effective temp. & $T_{eff}$   & 17,000 K    & Lennon et al. (\cite{Lennon-et-al-1993})\\
   Luminosity & $L_\star$ & $7\times10^5$ L$_\odot$ & Lennon et al. (\cite{Lennon-et-al-1993})\\
   Distance & d & 50,000 pc & Schaefer (\cite{Schaefer-2008})\\
\hline\hline
Fit parameters\\
\hline
Dust-shell parameters & Symbol & Value & Reference\\
\hline
   Inner dust radius & $R_{in}$ & $1.0\times10^4$ AU & -\\
   Outer dust radius & $R_{out}$ & $6.0\times10^4$ AU & -\\
   Dust mass & $M_{dust}$ & 0.01 M$_\odot$ & -\\
   Density distribution & $ \rho(r)$ & $ \rho(r) \approx r^{-2}$ & -\\
   Dust/gas & $f_{d/g}$  & 1/500 & van Loon et al. (\cite{vanLoon-et-al-1999}); Sargent et al. (\cite{Sargent-et-al-2011})\\
   Expansion velocity & $v_{exp}$ & 60 km/s & Wolf et al. (\cite{Wolf-et-al-1981})\\
\hline  
Dust composition & Symbol & Mass abundance & Reference\\
\hline
   Amorp. silicates & MgSiO$_3$ & 63\% & J\"{a}eger et al. (\cite{Jaeger-et-al-1994})\\ 
   Amorp. silicates & MgFeSiO$_3$ & 30\% & J\"{a}eger et al. (\cite{Jaeger-et-al-1994}\\
   Crys. silicates  & Mg$_2$SiO$_4$) & 3\% & J\"{a}eger et al. (\cite{Jaeger-et-al-1994})\\
   Metallic iron &Fe  & 3\%  & Pollack et al. (\cite{Pollack-et-al-1994})\\
   Polycyclic aromatic hydrocarbon & PAH  & 0.1\% & Draine \& Li (\cite{Draine-Li-2007})\\
\hline   
Dust grain properties & Symbol & Value & Reference\\
\hline
   Min. dust size (silicates) & $a_{min}$ & 0.01\,$\mu$m & Mathis, Rumple \& Nordsieck (\cite{Mathis-Rumple-Nordsieck-1977})\\
   Max. dust size (silicates) & $a_{max}$ & 0.1\,$\mu$m & Mathis, Rumple \& Nordsieck (\cite{Mathis-Rumple-Nordsieck-1977})\\
   Min. dust size (PAHs) & $a_{min}$ & 0.001\,$\mu$m & Mathis, Rumple \& Nordsieck (\cite{Mathis-Rumple-Nordsieck-1977})\\
   Max. dust size (PAHs) & $a_{max}$ & 0.0011\,$\mu$m & Mathis, Rumple \& Nordsieck (\cite{Mathis-Rumple-Nordsieck-1977})\\
\hline\hline
\end{tabular}
\end{table}
%

%
%



\end{document}